\documentclass[journal,twoside,web]{colors}
\usepackage{style1}
\usepackage{cite}
\usepackage{amsmath,amssymb,amsfonts}
\usepackage{algorithmic}
\usepackage{graphicx}
\usepackage{textcomp}
\usepackage{hyperref}
\usepackage{tikz}
\usepackage{textcomp}

\begin{document}
\title{Deformable Image Registration of Dark-Field Chest Radiographs for Local Lung Signal Change Assessment}

\author{Fabian Drexel, Vasiliki Sideri-Lampretsa, Henriette Bast, Alexander W. Marka, Thomas Koehler,\\ Florian T. Gassert, Daniela Pfeiffer, Daniel Rueckert, and Franz Pfeiffer
\thanks{This work was supported by the European Research Council (ERC Synergy Grant SmartX, SyG 101167328 and ERC AdG 884622), the Center for Advanced Laser Applications (CALA), the Federal Ministry of Education and Research (BMBF)
and the Free State of Bavaria under the Excellence Strategy of the Federal Government and the Länder, the
Technical University of Munich – Institute for Advanced Study.}
\thanks{F. Drexel and H. Bast are with the Chair of Biomedical Physics, Department of Physics, School of Natural Sciences, Technical University of Munich, Garching, Germany and the Munich Institute for Biomedical Engineering, Technical University of Munich, Garching, Germany, and the Department of Diagnostic and Interventional Radiology, School of Medicine and Klinikum rechts der Isar, Technical University of Munich, Munich, Germany. (email: fabian.drexel@tum.de; henriette.bast@tum.de).}
\thanks{V. Sideri-Lampretsa is with the Chair for AI in Healthcare and Medicine, Technical University of Munich (TUM) and TUM University Hospital, Munich, Germany. (e-mail: vasiliki.sideri-lampretsa@tum.de)}
\thanks{T. Koehler is with Philips Innovative Technologies, Hamburg, Germany, and with the Institute for Advanced Study, Technical University of Munich, Garching, Germany. (e-mail: thomas.koehler@philips.com)}
\thanks{A. W. Marka, F. T. Gassert, and D. Pfeiffer are with the Department of Diagnostic and Interventional Radiology, School of Medicine and Klinikum rechts der Isar, Technical University of Munich, Munich, Germany. D. Pfeiffer is also with the Institute for Advanced Study, Technical University of Munich, Garching, Germany. (e-mail: alexander.marka@tum.de; florian.gassert@tum.de; daniela.pfeiffer@tum.de)}
\thanks{D. Rueckert is with the Chair for AI in Healthcare and Medicine, Technical University of Munich (TUM) and TUM University Hospital, Munich, Germany, the Munich Center for Machine Learning (MCML), Munich, Germany and the Department of Computing, Imperial College London, UK. (e-mail: daniel.rueckert@tum.de)}
\thanks{F. Pfeiffer is with the Chair of Biomedical Physics, Department of Physics, School of Natural Sciences, Technical University of Munich, Garching, Germany, the Munich Institute for Biomedical Engineering, Technical University of Munich, Garching, Germany, Institute for Advanced Study, Technical University of Munich, Garching, Germany, and with the Department of Diagnostic and Interventional Radiology, School of Medicine and Klinikum rechts der Isar, Technical University of Munich, Munich, Germany (e-mail: franz.pfeiffer@tum.de).}}

\maketitle

\begin{abstract}
Dark-field radiography of the human chest has been demonstrated to have promising potential for the analysis of the lung microstructure and the diagnosis of respiratory diseases. However, previous studies of dark-field chest radiographs evaluated the lung signal only in the inspiratory breathing state. Our work aims to add a new perspective to these previous assessments by locally comparing dark-field lung information between different respiratory states.
To this end, we discuss suitable image registration methods for dark-field chest radiographs to enable consistent spatial alignment of the lung in distinct breathing states.
Utilizing full inspiration and expiration scans from a clinical chronic obstructive pulmonary disease study, we assess the performance of the proposed registration framework and outline applicable evaluation approaches.
Our regional characterization of lung dark-field signal changes between the breathing states provides a proof-of-principle that dynamic radiography-based lung function assessment approaches may benefit from considering registered dark-field images in addition to standard plain chest radiographs. 
\end{abstract}


\section{Introduction}
\label{sec:introduction}

\IEEEPARstart{D}{ark-field} radiography is an imaging technique that allows to visualize micro-structural features of an investigated sample \cite{b3,yashiro2010origin}. 
This is particularly promising in the field of lung imaging, as the alveoli form fine tissue-air structures that generate a considerable dark-field signal but hardly contribute to the signal in the attenuation domain due to the large air fraction. After demonstrating the application of the method for lung imaging in animals \cite{b4}, clinical studies investigated the potential for human lung assessments. The results from these studies demonstrated that dark-field chest X-ray imaging allows the visualization and detection of COVID-19-pneumonia \cite{b5} and the assessment and quantification of pulmonary emphysema in patients with chronic obstructive pulmonary disease (COPD) \cite{b6,b7}.

However, these previous evaluations on the human lung utilized the analysis of the dark-field signal in the inspiratory breathing state only.
Considering the substantial changes of the lung during the respiratory cycle and the impact of diseases on the breathing dynamics, comparing different respiratory states has the potential to expand the existing dark-field lung signal evaluation approaches toward functional assessments \cite{b4}. The challenge for such a comparative analysis is that a meaningful spatial correspondence between the images in the respiratory states is required. 

\begin{figure}[!t]
  \centering
  \includegraphics[width=\columnwidth]{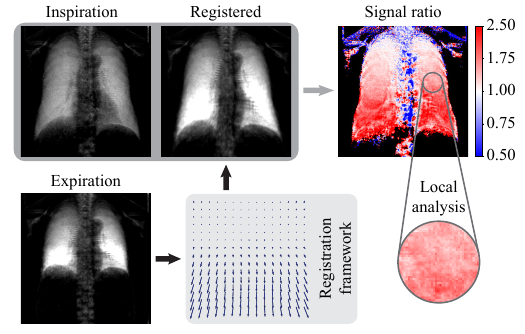}
  \caption{Dark-field-compatible deformable image registration methods enable matching chest radiographs in full expiration and full inspiration. 
  The ratio of the transformed expiration and inspiration dark-field images is used to analyze local lung signal changes.}
  \label{fig: Abstract_image}
\end{figure}

Therefore, this article aims to discuss suitable image registration methods specifically for dark-field chest radiographs to enable a local comparison of the lung signal in different phases of the respiratory cycle (see Fig.~\ref{fig: Abstract_image}). In addition, it is crucial for a comparative image analysis to employ suitable quantitative evaluation methods that can capture the characteristic dynamics of the signal changes properly. Inspired by the methodology used in dynamic chest radiography (DCR)~\cite{b13}, we explore possible analysis approaches and discuss the differences between the dark-field and commonly utilized attenuation signal~domains.
\noindent \\
In summary, the main contributions of our work are as follows:

1) We propose an iterative optimization-based deformable registration framework that enables the matching of dark-field chest radiographs acquired in different respiratory states. Since dark-field chest radiography is still a relatively new imaging method compared to established techniques, the framework is the first to be specifically tailored for this registration task. 

2) We investigate the performance of the proposed registration framework considering full inspiration and expiration dark-field scans from a clinical COPD study and discuss registration evaluation metrics for this image-matching task.

3) Considering the registered images, our local characterization of the lung dark-field signal changes between respiratory states provides promising indicators that radiography-based lung function assessment approaches may benefit from utilizing the dark-field information in addition to the attenuation signal.

\section{Related Work}

Image registration has evolved into an indispensable and widespread method in the clinical practice for accurately combining information from different image domains~\cite{b8,b20}. 
Because of the respiratory dynamics and variability, image registration is particularly important in lung imaging to reliably align data from different time points~\cite{b10}, patients~\cite{li2003establishing}, or modalities \cite{pan2018fast}.
The registered images can then be leveraged for diagnostics \cite{li2014deformable}, disease progression monitoring~\cite{stavropoulou2021multichannel}, and treatment planning purposes \cite{b11}.

When it comes to the comparative analysis of lung image information throughout the respiratory cycle, image registration is a valuable tool specifically with regard to functional examinations \cite{b10,b13.1}.
For serial thoracic computed tomography (CT) scans, registration algorithms have been utilized within intensity-based and Jacobian-based methods that aim to provide an easily accessible and cost-effective alternative for lung ventilation assessments to other standard techniques, such as positron emission tomography or Xenon-enhanced CT~\cite{ding2010comparison}.
However, despite the wide application of thoracic CT image registration methods, it remains a very challenging task due to the vast three-dimensional structural complexity and the large and anisotropic deformations \cite{xiao2023deep}.
Furthermore, as a series of CT images is required for the above-mentioned registration-based ventilation analysis, the patient is exposed to a comparably high radiation dose.

DCR aims to mitigate this issue by utilizing registered sequential plain chest radiographs instead of CT scans for functional examinations \cite{b13.1}.
Moreover, in contrast to standard CT, DCR can be performed in a sitting or standing patient position, reflecting physiologically relevant daily activity \cite{b13}.
To analyze pulmonary ventilation, DCR studies generally calculate X-ray image pixel value change rates or differences~\cite{b22,b23}.
However, in order to determine the local ventilation from the slight dynamic changes throughout the respiratory cycle, several additional image processing steps are necessary to separate the relevant variations from the influences of bone overlay in the lung area and blood flow changes \cite{b13}.

\section{Methods}

\begin{figure*}[!t]
  \centering
  \includegraphics[width=\textwidth]{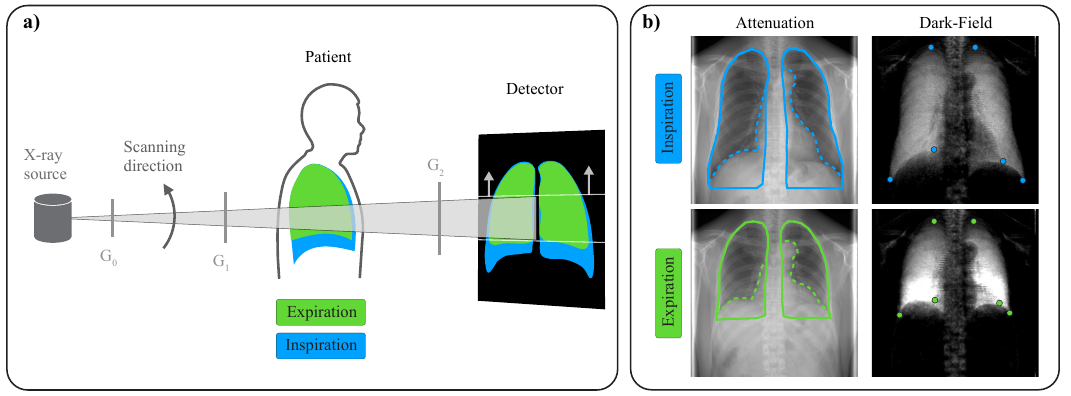}
  \caption{\textbf{a)} Schematic view of the dark-field chest X-ray prototype system with the main components. The dark-field lung image formation in the expiratory and inspiratory breath-hold states is indicated in green and blue. \textbf{b)} Examples of the corresponding attenuation and dark-field images with an overlay of the manually drawn full (solid line) and partial (dotted line) lung masks and the annotated landmarks (blue and green dots).}
  \label{fig: DF scanner}
\end{figure*}

\subsection{Dark-Field Chest Radiography}

\subsubsection{Dark-field chest X-ray prototype system} The dark-field radiographs studied in this article were acquired at a clinical dark-field prototype system which is located at the university hospital Klinikum rechts der Isar, Technical University of Munich, Germany, and uses grating-based X-ray interferometry~\cite{b3}. 
A schematic view of the set-up is shown in Fig.~\ref{fig: DF scanner}\textbf{a}.

The Talbot-Lau-type interferometer consists of a source grating $G_0$ to provide spatially coherent X-rays \cite{b14}, the intensity pattern inducing grating $G_1$, and the analyzer grating $G_2$ in front of the detector (effective pixel size of 444 $\mu$m $\times$ 444 $\mu$m). 
For details regarding the X-ray tube, gratings, and the detector, refer to \cite{b19}. 
As the fabrication of gratings large enough to cover the full desired field of view was not feasible at the time of construction, the radiography system works in scanning mode. This means that during image acquisition, the movable interferometer plane with the gratings scans the field of view from bottom to top within 6.5 seconds (as indicated in Fig. \ref{fig: DF scanner}\textbf{a}). 

Following the scanning procedure, the attenuation and dark-field signals can be reconstructed simultaneously from the acquired data \cite{b17,b18,b19}. 
As this method provides attenuation and dark-field signals based on the same data recording, the two reconstructed images are perfectly co-registered.

\subsubsection{Dataset} For our analyses, we used image data from a clinical study conducted among participants with or without COPD.
The study was approved by the Institutional Ethics Committee of the Technical University of Munich and by the national Radiation Protection Agency (Z5–22462/2–2017-021). All participants of this study gave written informed consent.
Participants with signs of emphysematous impairment or healthy lungs according to a CT evaluation were included. Participants with pulmonary pathologies or conditions other than COPD were excluded from the study. For further details regarding the study, refer to \cite{b6}. 
Within the study imaging protocol, attenuation and dark-field images were acquired at maximum inspiration and full expiration in the lateral and posteroanterior (PA) orientations. PA image examples are depicted in Fig. \ref{fig: DF scanner}\textbf{b}. The dataset includes the respective images of a total of 95 participants.

\subsection{Image Registration}

To establish the spatial correspondence between the PA attenuation or dark-field images in the two respiratory states, we utilized image registration methods.

For the formal definition of the image registration task, we adopted the common description as an optimization problem between a fixed $\mathcal{F}$ and a moving image $\mathcal{M}$ that is to be transformed to the fixed image \cite{b20}.
Within this setting, the generic objective function for the transformation optimization consists of a similarity term $\mathcal{L}_{\mathrm{sim}}$ and a regularization term~$\mathcal{L}_{\mathrm{reg}}$:
\begin{equation}
    \mathcal{L(\phi,\mathcal{M},\mathcal{F})} = \mathcal{L}_{\mathrm{sim}}(\mathcal{M}  \circ \phi , \mathcal{F}) + \alpha \mathcal{L}_{\mathrm{reg}}(\mathcal{\phi}).
    \label{eq: objective_function}
\end{equation}
Within this objective function, the similarity term quantifies the level of alignment between the fixed image $\mathcal{F}$ and the transformed moving image $\mathcal{M}  \circ \phi$.
The term $\mathcal{L}_{\mathrm{reg}}$ regularizes the transformation $\phi$ and may be seen as a way to introduce prior knowledge regarding the transformation solution into the objective function \cite{b20}. 
The regularization parameter $\alpha$ is used to adjust the relative influence of the regularization term to the similarity term within the objective function.
Following this generic formulation, selecting the specific optimization method, objective function terms, and the deformation model for the transformation generally depends on the particular image registration task. The selections for our dark-field chest radiograph registration framework will be introduced and explained in the following.

\subsubsection{Image preprocessing}
Within the first step of the preprocessing, the study participant images were padded with zeros from their initial shape of 956 $\times$ 947 to 956 $\times$ 956 pixels. This simplifies the pixel distance calculations for downsampled and interpolated images, which is relevant for the evaluation metrics. 
Additionally, we manually corrected horizontal shifts between the inspiration and expiration scan for some study participants where a major lateral displacement was evident. The lateral shifts can occur if study participants reposition themselves within the setup cabin between the inspiration and expiration image acquisition. 
Furthermore, for shorter computation times, the padded and shift-corrected images were downsampled to 256 $\times$ 256 pixels, resulting in effective pixel dimensions of 1.66~mm $\times$ 1.66~mm.

\subsubsection{Dark-field radiograph registration} 
For the registration task, we defined the inspiration radiograph as the fixed ($\mathcal{F}$) and the expiration scan as the moving image ($\mathcal{M}$).

Due to the expected large and complex lung deformations between the full inspiratory and expiratory breathing state, we opted for a multiresolution strategy, which can handle local traps within the optimization space in such registration tasks \cite{b26}. 
Within this approach, the downsampled input dark-field images are further downsampled and smoothed to reduced complexity levels, and the degree of freedom of the transformation model is limited accordingly. 

Regarding the non-rigid transformation model for the dark-field registration framework, we adopted a diffeomorphic spatial transformation approach using stationary velocity free-form deformations (SVFFD) \cite{b31}.
This approach was selected because the stationary velocity ansatz, by design, leads to consistent invertible transformations that preserve image topology, while the appealing aspects of the conventional free-form deformation like its computational efficiency are retained \cite{b27}. 

For the similarity term $\mathcal{L}_{\mathrm{sim}}$ within the objective function for the dark-field image registration task, we selected local normalized cross-correlation (LNCC) \cite{b29} over simpler and faster statistical similarity measures like the sum of squared differences (SSD) or the standard normalized cross-correlation (NCC). 
\begin{figure*}[!t]
  \centering
  \includegraphics[width=\textwidth]{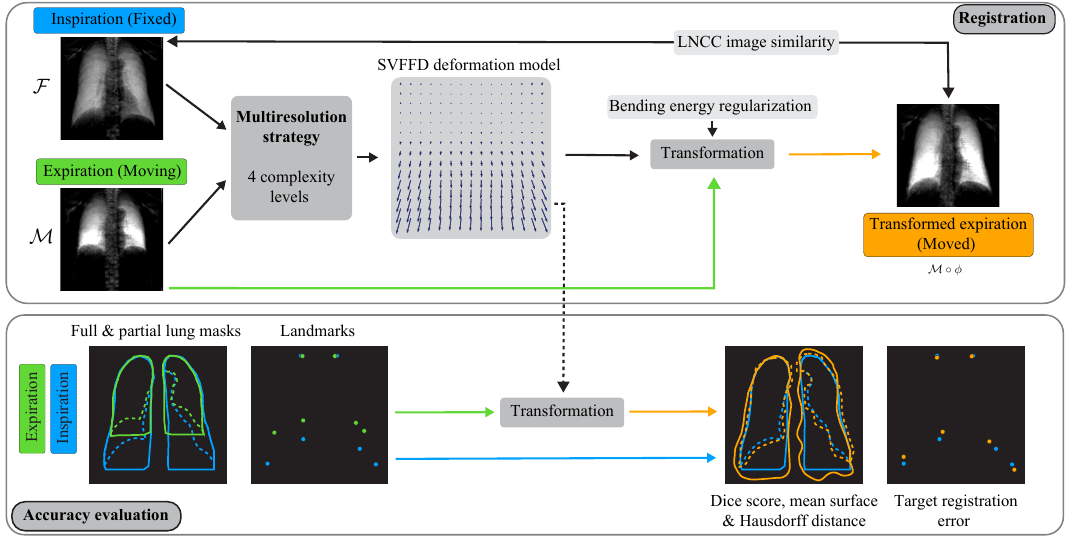}
  \caption{Overview of the dark-field chest radiograph registration framework (top) and the accuracy evaluation approaches (bottom). The expiration dark-field image is registered to the inspiration dark-field radiograph utilizing a multiresolution strategy and an SVFFD deformation model. The LNCC image similarity between the inspiration and transformed expiration image as well as the bending energy regularization guide the transformation optimization. Manually annotated full and partial lung masks as well as landmarks are utilized for the registration accuracy evaluation. The registration transformation is applied to the respective masks and landmarks. The Dice score, the mean surface distance, the Hausdorff distance of the masks, and the target registration error of the landmarks are calculated to assess the registration accuracy.}
  \label{fig: Reg_framework}
\end{figure*}
This is because the underlying assumption of SSD that the difference between the registered images should be zero except for noise \cite{b30}, is not a sensible approach due to the potentially prominent dark-field signal differences between the expiration and inspiration image (see e.g. Fig. \ref{fig: DF scanner}\textbf{b}) that should be preserved during the registration procedure.
NCC is based on the more general assumption that there is a linear relationship between the two images on the global image domain scale \cite{b30}. Extending the NCC assumption to a local image domain perspective, the LNCC similarity measure enables the registration framework to capture potentially different regional relationships between the dark-field signal in distinct lung~regions. 

From an anatomical point of view, the movement of the lung tissue in the two-dimensional projection plane should be continuous and without sliding or folding phenomena. Note that the sliding motion between rib cage and lung is irrelevant here since the rib cage does not generate a prominent dark-field signal. This prior knowledge was included via a bending energy regularization term in the objective function to ensure the smoothness of the transformation.
The bending energy is a popular penalty term for non-affine transformations \cite{b28}.\\
As a compact overview, the combination of the selected components and the general workflow within the registration framework is depicted in the upper part of Fig. \ref{fig: Reg_framework}.

\subsubsection{Registration evaluation}
As there is no ground truth transformation for our registration task, we used surrogate measures to assess the registration accuracy and regularity.
For the accuracy evaluation, we utilized mask-based and landmark-based measures, see the lower part of Fig. \ref{fig: Reg_framework}.

For the latter, a radiologist (co-author A. W. Marka) with previous experience with dark-field chest radiography annotated six landmarks to the inspiration and expiration dark-field images respectively, see Fig. \ref{fig: DF scanner}\textbf{b}. 
The annotated landmarks refer to points on the lung apex and on the costophrenic and costocardiac angles on the right and left sides of the lung. The respective landmarks were then transformed with the final registration transformation and the mean target registration error (TRE) was calculated \cite{b8}. As the TRE is based on Euclidian distances between the landmarks, it has immediate physical meaning in how accurately the characteristic points in the lung match after registration.

Since the landmarks are sparsely distributed we also assessed the accuracy via the mask-based measures. 
The full and partial lung masks (see solid/dashed green and blue lines in Fig. \ref{fig: DF scanner}\textbf{b}) were manually drawn onto the attenuation images.
Since the attenuation and dark-field images are perfectly co-registered within this imaging method, the masks generated using attenuation images also match the lung area in the corresponding dark-field domain.
The full masks include the heart region and the areas below the hemidiaphragm whereas the partial masks exclude these regions. 
For the mask-based accuracy metrics, we chose the Dice score (DICE), symmetrical mean surface distance (MSD), and the symmetric Hausdorff distance (HD) \cite{devos2019deep}. DICE measures the overlap of the lung mask of the inspiration image and the transformed expiration lung mask whereas MSD and HD quantify the distance between the outlines of the masks. As the surface distance measures are based on Euclidian distances of mask surface points, they provide physical interpretability regarding how well the lung masks match after registration.

The transformation regularity, on the other side, was evaluated based on the determinant of the  Jacobian $\text{det}(J) = \text{det}( \nabla \phi )$ \cite{devos2019deep}.
Similar to \cite{b31}, we calculated the folding ratio, which is the fraction of pixels in the image domain where $\text{det}(J) < 0$ (locations where the image topology is not preserved). In addition, we computed the mean magnitude of the gradient of the Jacobian determinant (MMGJD), which is an indicator of the spatial smoothness of the transformation~\cite{b31}.

\subsubsection{Comparison methods} 
To be able to compare the dark-field image analysis with standard plain chest radiograph evaluations, we also implemented a framework for the registration of the corresponding attenuation images.
For this purpose, we utilized a modified version of the dark-field image registration framework described above.

As a first modification step, we opted for an additional affine pre-registration procedure before the non-rigid registration. This approach aimed to facilitate the local non-rigid deformations with the increased information content of the attenuation images (bone/soft tissue signal in the lung region) by using an affine registration to establish a rough spatial correspondence in advance. The objective function for the affine pre-registration procedure consists only of a masked NCC loss similarity term.
The masking was necessary to enable a reasonable registration of the lung regions by reducing the impact of the collimator attenuation signal in some images and the extensive additional information content outside the lung region due to the bone and soft tissue structures.
For the masking, we used the previously described partial lung mask for the inspiration image, dilated them by 25 pixels, and then computed the concave hull of the dilated masks (\emph{concave\_hull} Python package with concavity 300, see e.g. \cite{park2012concave}).

As a second modification step, we selected a masked LNCC loss within the non-rigid multiresolution strategy transformation optimization for the same reasons as described in the context of the affine registration. The affinely pre-registered expiration attenuation image was used as the moving image input for the non-rigid registration procedure.

\subsubsection{Implementation details} 
To implement and test the previously described strategies and methods, we built upon the functionality of Deepali \cite{b21}, a Python library that enables both iterative optimization-based and learning-based registration. 
Due to the limited data availability, only the iterative optimization functionalities and no learning-based approaches were explored. 
Deepali provides a variety of different deformation models as well as similarity and regularization terms for the customization of a suitable objective function.
Unless otherwise specified below, the corresponding functions and models from Deepali have been used within our registration frameworks.
Deepali is built on the popular PyTorch machine-learning library, allowing for fast registration optimization using graphics processing units (GPUs). This is also the main reason why we decided to use Deepali instead of other commonly used registration libraries that are still mainly reliant on CPU-based optimization. 
In our case, all registration optimizations were conducted on \emph{NVIDIA GeForce GTX 1080 Ti} GPUs.

Within the multiresolution strategy, we used four complexity levels corresponding to image dimension sizes of 256, 128, 64, and 32 pixels.
A Gaussian kernel was used for smoothing with the default standard deviation of the respective function within Deepali.
After the initialization of the four complexity levels, the transformation was optimized on the lowest-resolution level first, until convergence (adaptive optimization) or the maximum number of optimization steps (3500 for the non-rigid case) is reached.
The convergence criterion within the adaptive optimization was that the absolute value of the difference between the current objective function value and the moving average within a window of the last 200 steps is lower than the defined convergence threshold ($10^{-3}$ times the initial objective function value). 
After the transformation optimization on a specific level, the same procedure was continued on the next complexity level until the input resolution was reached.  
The PyTorch Adam optimizer was used for all registration transformation optimizations.

The specific parameters used for the dark-field and attenuation image registration algorithms are listed in Table \ref{tab: implementation_parameters}. 

\begin{table}[h!]
    \caption{Registration Framework Parameters}
    \centering
    \renewcommand{\arraystretch}{1.2} 
    \begin{tabular}{|p{50pt}|p{60pt}|p{50pt}|p{40pt}|} 
        \hline
         \textbf{Parameter} & \textbf{Non-rigid dark-field} & \textbf{Affine attenuation} & \textbf{Non-rigid attenuation}  \\
        \hline
        Stride (px) & (8, 9, 10) & - & 10  \\
        \hline
        LNCC kernel sizes (px) & 11, 21, 41, 81 & - & 5, 11, 21, 41  \\
        \hline
        Optimization steps & adaptive & 800, 50, 50, 50 & adaptive  \\
        \hline
        Learning rate & $10^{-4}$ & $10^{-3}$ & $10^{-4}$  \\
        \hline
        $\alpha$ & (1, 100, 200, 500) & - & 80  \\
        \hline
        \multicolumn{4}{p{240pt}}{Overview of the selected parameters for the affine/non-rigid registration implementations for the attenuation and dark-field images. Listed numbers within brackets indicate varying parameters. Listed numbers without brackets indicate the parameter with increasing resolution level.}
    \end{tabular}
    \label{tab: implementation_parameters}
\end{table}
\noindent As the non-rigid transformations within the SVFFD approach are parameterized by cubic B-spline functions \cite{b28}, the distance between control points (stride) on the image pixel lattice needs to be provided for the transformation initialization. 
Since an appropriate stride-regularization weighting factor $\alpha$ combination for the transformation model within the multiresolution approach is difficult to infer a priori, we evaluated the registration results for certain sets of values (see Table \ref{tab: implementation_parameters}). 
The same stride and $\alpha$ were used on all the resolution levels.

Furthermore, as the LNCC function calculates the NCC in local quadratic windows, suitable kernel sizes need to be provided for the registration parts that utilize LNCC (see Table \ref{tab: implementation_parameters}). 
The kernel sizes for the dark-field registration framework were chosen so that the calculation window scale corresponds approximately to the width of one side of the lung in a dark-field scan in the inspiratory breathing state at the corresponding resolution level. This aims to be able to capture different relationships on this scale. For the non-rigid attenuation image registration part, the kernel sizes were reduced to allow capturing different relations on smaller scales that may be caused by the overlapping bone and soft tissue structures in the lung area.

For the mask-based accuracy assessments, the publicly available code implementations from \cite{b31} were used to calculate the Dice score, MSD, and HD.

\subsection{Analysis of the Registered Images}
In the context of dark-field signal change quantification, we drew inspiration from earlier works that addressed analogous issues and followed a similar approach as in DCR.
However, instead of the commonly used inter-frame difference ansatz~\cite{b13}, we opted for an inter-frame ratio approach. 
As the dark-field signal can be expressed via the product of the dark-field coefficient $\epsilon$ (representing the sample microstructure dependence) and the sample thickness $d$ on the image pixel-associated beam path as $D = \epsilon \cdot d$ \cite{bech2010quantitative}, we can form the ratio of the registered expiration and the inspiration dark-field signal (for every image pixel):
\begin{equation}
    R_{D}=\frac{D_{\mathrm{exp.},\mathrm{reg.}}}{D_{\mathrm{insp.}}}=\frac{\epsilon_{\mathrm{exp.},\mathrm{reg.}}}{\epsilon_{\mathrm{insp.}}} \cdot \frac{d_{\mathrm{exp.},\mathrm{reg.}}}{d_{\mathrm{insp.}}}.
    \label{eq: DF_signal_ratio}
\end{equation}
\noindent In contrast to a signal difference ansatz, this ratio approach allows us theoretically to separate the relative dark-field coefficient and lung thickness variation contributions to the dark-field signal change (two fraction terms in \eqref{eq: DF_signal_ratio}).  
The same evaluation method was used for the attenuation radiographs to compare both the attenuation ratio $R_{A}$ and the dark-field ratio $R_{D}$ characteristics within the analyses.

We assessed the signal ratios with two approaches. 
Inspired by evaluations in DCR \cite{b22}, we also projected the left and right lung signal ratios within the partial masks onto the craniocaudal (CC) axis. With this approach, we reduce the two-dimensional signal ratio information to a simpler mean ratio graph which depends on the distance from the lung apex. The complexity is then further reduced to a scalar value by generating a linear fit for the graph to characterize signal ratio changes from upper to lower lung regions via the slope (CC gradient) of the fit. To fit a linear function to the mean signal ratio data, we utilized the \emph{HuberRegressor} function from the Python \emph{scikit-learn} library.

For the alternative dark-field signal assessment approach, we divided the partial lung masks into an upper, middle, and lower region, see Fig. \ref{fig: Mean_DF_signal_change}, right. 
Then, we selected the signal ratio values within the respective region focus and calculated the mean dark-field ratios in the three lung fields. 

The evaluation results for the study participants were then further investigated concerning breathing capacity during imaging and COPD severity. 
The breathing capacity was quantified via the normalized relative vital lung capacity of the study participants: $VLC_{\mathrm{rel}}=(V_{\mathrm{insp.}}-V_{\mathrm{exp.}})/V_{\mathrm{insp.}}$. 
For the $VLC_{\mathrm{rel}}$ calculation, the inspiration and expiration lung volumes $V_{\mathrm{insp.}}$, $V_{\mathrm{exp.}}$ were estimated utilizing the lateral and PA attenuation radiographs following the approach in \cite{b32}. 
The relationship between CC gradients or mean signal ratios and the breathing capacity were then assessed considering the Spearman's correlation coefficient (\emph{SciPy} Python package) with $VLC_{\mathrm{rel}}$.

For the COPD severity quantification, we used Fleischner Society emphysema classification scores for the study participants, graded by trained radiologists after visual assessments of CT images, see \cite{b6}. Within this scheme, the study participants were divided into the following emphysema groups: 0~-~absent, 1 - trace, 2 - mild, 3 - moderate, 4 - confluent, and 5~-~advanced destructive.

\section{Results and Discussion}

\subsection{Registration of Dark-Field Radiographs}
For the dark-field registration experiments, we excluded the study participants where parts of the lung were out of the image domain, resulting in a total number of 87 image pairs.
\begin{table}[h!]
    \caption{Framework Parameter Registration Impact}
    \centering
    \renewcommand{\arraystretch}{1.2} 
    \begin{tabular}{|p{24pt}|p{18pt}|p{18pt}|p{18pt}|p{18pt}|p{16pt}|p{18pt}|p{16pt}|} 
        \hline
        \textbf{$\alpha$, Stride} & \textbf{DICE full} & \textbf{DICE part.} & \textbf{MSD full (mm)} & \textbf{MSD part. (mm)} & \textbf{TRE (mm)} & \textbf{MMG JD $\cdot 10^{-3}$} & \textbf{Dura- tion (s)} \\
        \hline
        1, 8 & 0.943 & 0.936 & 4.33 & 3.57 & 5.97 & 37 & 157 \\
        \hline
        1, 9 & 0.944 & 0.937 & 4.27 & 3.49 & 5.81 & 28 & 159 \\
        \hline
        1, 10 & 0.944 & 0.937 & 4.12 & 3.45 & 5.87 & 23 & 156\\
        \hline
        100, 8 & 0.949 & 0.937 & 3.81 & 3.46 & 5.85 & 6 & 105\\
        \hline
        100, 9 & 0.951 & 0.936 & 3.78 & 3.49 & 5.95 & 5 & 127\\
        \hline
        100, 10 & 0.950 & 0.936 & 3.71 & 3.52 & 6.10 & 4 & 103\\
        \hline
        200, 8 & 0.950 & 0.936 & 3.74 & 3.49 & 6.04 & 4 & 99\\
        \hline
        200, 9 & 0.951 & 0.936 & 3.72 & 3.51 & 6.26 & 4 & 118\\
        \hline
        200, 10 & 0.951 & 0.935 & 3.69 & 3.59 & 6.36 & 3 & 101\\
        \hline
        500, 8 & 0.951 & 0.935 & 3.68 & 3.58 & 6.32 & 3 & 104\\
        \hline
        500, 9 & 0.952 & 0.934 & 3.65 & 3.73 & 6.55 & 3 & 121\\
        \hline
        500, 10 & 0.952 & 0.933 & 3.62 & 3.81 & 6.80 & 2 & 112\\
        \hline
        \multicolumn{8}{p{240pt}}{Impact of the regularization parameter $\alpha$ and the transformation model stride on the median of evaluation metrics and the optimization duration. full = full lung masks; part. = partial lung masks; DICE = Dice score; TRE = target registration error; MSD = mean surface distance; MMGJD = Jacobian determinant mean gradient magnitude.}
    \end{tabular}
    \label{tab: param_gridsearch}
\end{table}

\subsubsection{Regularization weighting and stride parameter impact}
We explored the impact on the registration result within the $\alpha$-stride parameter space via multiple evaluation metrics. 
The impact on the medians of the Dice score and the MSD for the full and partial lung masks, as well as the influence on the TRE, MMGJD, and the registration duration, are summarized in Table \ref{tab: param_gridsearch}.
Median values were evaluated to capture the behavior of potentially skewed distributions with outliers.

The median Dice score for the full lung mask shows slightly improved values for increasing regularization weighting and stride. The median Dice score for the partial lung mask exhibits the opposite trend.
The median MSD for the full lung mask shows better values for increasing regularization strength and stride. The partial mask median MSD first improves for increasing regularization weighting and stride but then deteriorates after $\alpha=100$. The same holds for the median TRE.
Within the explored parameter space, regularization weighting of $\alpha=1$ leads to the highest median MMGJD values and registration durations.
\begin{figure*}[!t]
  \centering
  \includegraphics[width=\textwidth]{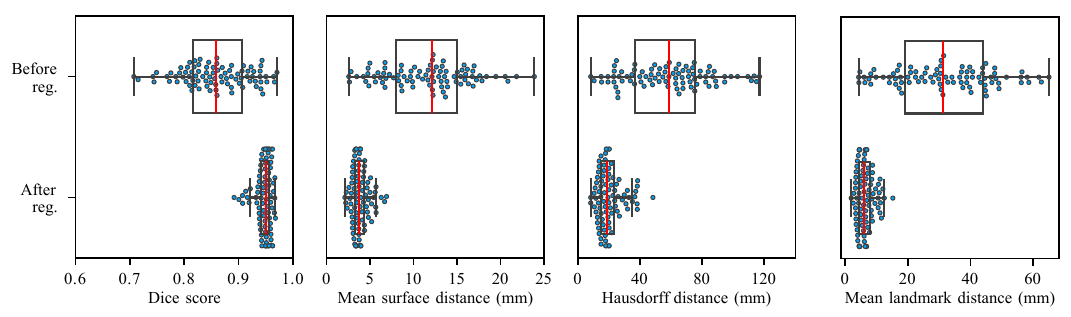}
  \caption{Registration evaluation metrics considering full lung mask metrics and landmark distances (TRE) for the dark-field image registration framework within the dataset. The top row shows the scores before, and the bottom row the values after the registration procedure. Red lines indicate the median, black boxes the interquartile range (25th to 75th percentile), and blue dots the individual study participant scores.}
  \label{fig: Reg_performance}
\end{figure*}
\noindent This behavior is expected as the bending energy loss imposes fewer restrictions and more less-smooth transformation states are accessible in the optimization search space for weak regularization.
Furthermore, for the configuration ($\alpha=1$, stride = 8), a non-zero folding ratio in the order of $10^{-4}$ was observed for two cases.
This indicates that such low regularization strength leads to violations of topology preservation even for the SVFFD model where we would not expect folding due to the velocity field formulation.

In summary, in contrast to the prominent changes for the median MMGJD and registration duration, the other metric medians barely change considerably within the evaluated $\alpha$-stride parameter space. 
This is because the effective pixel dimensions for the downsampled input images are 1.66 mm whereas the magnitude of MSD and TRE variations is in the order of 1 mm.
Nevertheless, if we consider the metric behaviors described above, the opposing trends make it impossible to select the optimal configuration within the explored search space.
Only the image domain folding and less smooth transformations for $\alpha=1$ revealed that these parameter configurations are not suitable for our task. Apart from that, the configurations generally provide high Dice and low surface distance and TRE scores, considering that uncertainties due to the manual annotation of the lung masks and landmarks are to be expected.
Considering all findings, we selected the model configuration ($\alpha=100$, stride = 10) for further registration experiments.

\subsubsection{Selected configuration performance details} 
Based on the selected parameter configuration, the performance of the registration framework on the dataset was investigated in more detail. Fig. \ref{fig: Reg_performance} shows a comparison of the full mask metric score and TRE distributions before (upper row) and after the registration procedure (bottom row). 

Considering the distribution medians, we observed a pronounced increase in the Dice score from 0.858 to 0.950, an MSD reduction from 12.1~mm to 3.7~mm, and a substantial drop in the HD from 58.9~mm to 18.5~mm. Similar behavior was observed for the partial masks with median DICE shifting from 0.835 to 0.936, a median MSD reduction from 10.4~mm to 3.5~mm, and an HD decrease from 63.1~mm to 18.9~mm.
The analogous response of the full and partial mask-based metrics is not surprising, as they are identical in many regions. Nevertheless, the assessment with the full and partial masks is important to inspect the behavior of the registration below and at the diaphragmatic arches as well as at the borders to the heart area.
A substantial improvement of the median from 31.2~mm to 6.1~mm is also recognizable for the mean distances of the landmarks (TRE).
The landmark evaluation approach has the advantage over the surface distance metrics that the performance can be examined at specific characteristic points of the lung in the dark-field domain instead of an entire lung region surface.
One general limitation of the metrics used, however, is that the landmarks and masks utilized can be used to evaluate the accuracy of registration mainly at the boundaries of the lungs to the thorax, diaphragm, or heart, but not at central points within the lung.

\begin{figure*}[!t]
  \centering
  \includegraphics[width=\textwidth]{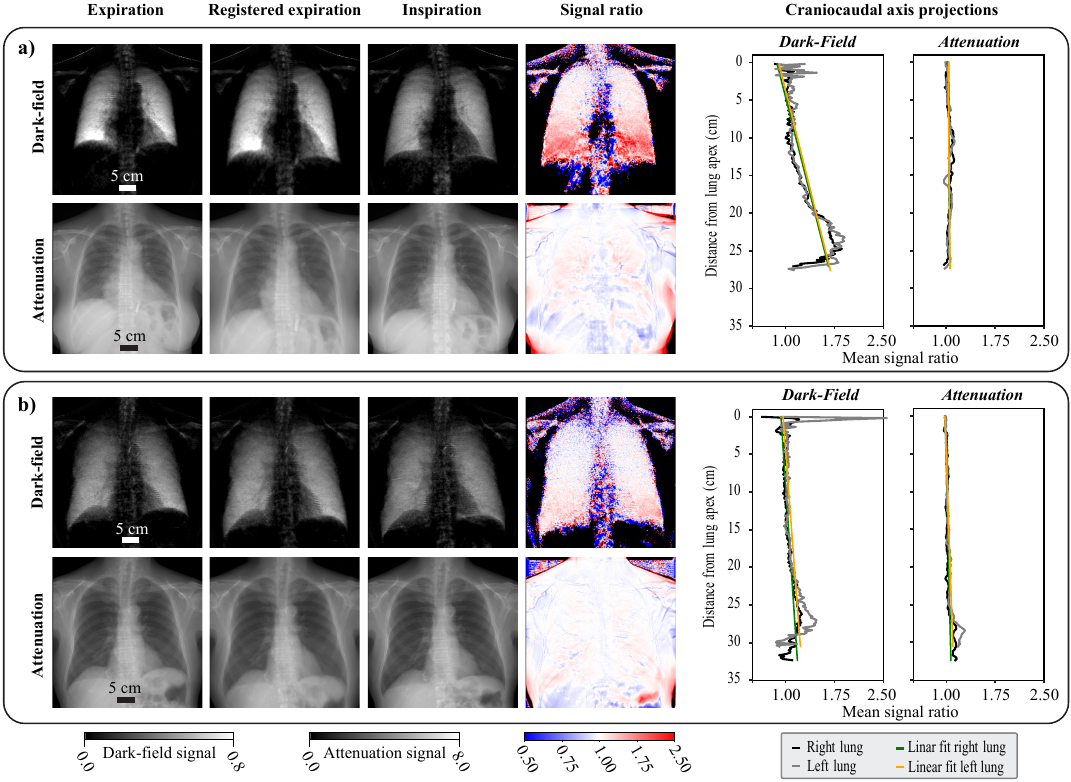}
  \caption{Attenuation and dark-field image registration results for two study participants with signal ratio analysis. \textbf{a)} Fleischner scale score 0, $VLC_{\mathrm{rel}}=0.28$. \textbf{b)} Fleischner scale score 2, $VLC_{\mathrm{rel}}=0.14$. The linear fit slopes to the CC signal ratio projection differ substantially for the two study participant cases in the dark-field domain (0.029 right and left for case \textbf{a)}, 0.008 right and 0.009 left for case \textbf{b)}) in contrast to the attenuation domain (0.001 right and left for case \textbf{a)}, 0.003 right and 0.004 left for case \textbf{b)}). Black regions within the ratio images indicate pixels where the transformed exp. or insp. signal was below $10^{-15}$ and thus set to \emph{NaN} to avoid division by zero errors.}
  \label{fig: Reg_examples}
\end{figure*}

Overall, despite the considerable variance of the initial configuration of the lung in expiration and inspiration (broad distributions in the top row of Fig.~\ref{fig: Reg_performance}), the registration framework demonstrates consistently good performance for all cases (narrow distributions in the bottom row of Fig.~\ref{fig: Reg_performance}). In line with the quantitative evaluation metrics, the visual qualitative evaluation confirmed the good performance of our dark-field registration framework within the dataset (see e.g. Fig.~\ref{fig: Reg_examples}\textbf{a}, \textbf{b}).

\subsection{Analysis of the Registered Images}
The successfully registered images were then utilized for the evaluation of the transformed expiration-inspiration dark-field signal ratios. Two examples of the dark-field ratio $R_{D}$ evaluations are depicted alongside the initial and registered image in the top row of Fig.~\ref{fig: Reg_examples}\textbf{a}, \textbf{b}. 
Other structures than the lung such as the clavicle bones or the spine are also visible within the signal ratio images. 
However, the focus is on the signal ratio in the lungs, which is why the quantitative analyses were carried out using lung masks.

\subsubsection{Dark-field and attenuation domain comparison} To illustrate the differences between the signal change in the lung from inspiration to expiration in the dark-field and attenuation domains, the registered attenuation images with the corresponding signal ratios $R_{A}$ are displayed in the bottom row of Fig.~\ref{fig: Reg_examples}\textbf{a}, \textbf{b}. 
These two study participants exemplify differences in signal changes primarily related to the breathing capacity during the image acquisition.
The study participant in Fig.~\ref{fig: Reg_examples}\textbf{a} represents the higher breathing capacity cases, indicated by $VLC_{\mathrm{rel}}=0.28$. 
In contrast, the study participant in Fig.~\ref{fig: Reg_examples}\textbf{b} represents the lower breathing capacity cases, indicated by $VLC_{\mathrm{rel}}=0.14$. 

Comparing the two signal domains, more distinct signal ratio differences between the cases are recognizable in the dark-field than in the attenuation domain. 
This is partly because the magnitude of the changes in the lung area is larger in the dark-field signal. 
But more importantly, there is signal overlap from different objects in the attenuation domain. 
In particular, streaks are recognizable in the lung area of the attenuation signal ratio in many areas caused by the clavicles and ribs. 
As these bone structures don't follow the movement of the lung tissue, the registration algorithm for the attenuation domain is unable to correctly map these structures to each other when correcting for lung tissue motion.
As these bone structures only generate a very weak dark-field signal compared to lung tissue, this issue is overcome to a large extent with the registered dark-field radiographs.
In the attenuation domain, this could only be solved by additional image processing steps such as bone suppression algorithms. 
The same signal overlap aspect also applies to non-lung soft tissue structures such as breast fat tissue, which generates a substantial attenuation signal and can superimpose the lung area in considerably different locations in the expiratory and inspiratory state.

To quantitatively assess the differences in signal ratios between the dark-field and attenuation domains, we evaluated the CC axis gradients, see right side of Fig.~\ref{fig: Reg_examples}\textbf{a}, \textbf{b}. 
For the example cases, prominent slope differences are observed within the dark-field domain in contrast to the attenuation image domain. 
This indicates that signal change differences connected with breathing capacity can be analyzed more easily and more accurately by comparing the dark-field signal in different respiratory states.
To substantiate this hypothesis derived from the single study participant pair, we compared the CC axis gradient about $VLC_{\mathrm{rel}}$ for 53 study participants (attenuation registration metric medians: full/part. mask DICE 0.943/0.924, MSD 3.82/3.91~mm, HD 19.90/20.24~mm). 
The other study participants were excluded from this evaluation due to insufficient attenuation image registration quality. 
We found stronger correlations of the dark-field-based slopes (Spearman corr.\ coeff.: $r_{s,right}=0.55$, $p<0.01$ and $r_{s,left}=0.48$, $p<0.01$) compared to the attenuation-based CC gradients (Spearman corr.\ coeff.: $r_{s,right}=0.20$, $p=0.16$ and $r_{s,left}=0.40$, $p<0.01$). The correlation of the attenuation-based CC slope on the right side of the lung is even not significant considering the utilized significance threshold of $p=0.05$.
In addition to the stronger correlations, the dark-field-based slopes change over a broader range (dark-field: $\approx -0.02$ to $0.06$, attenuation: $\approx 0.00$ to $0.02$). Combined, this indicates a general trend that the CC gradient in the dark-field domain shows more pronounced changes with the breathing capacity than in the attenuation domain.

\subsubsection{Lung field signal characterization} 
\begin{figure}[!t]
  \centering
  \includegraphics[width=\columnwidth]{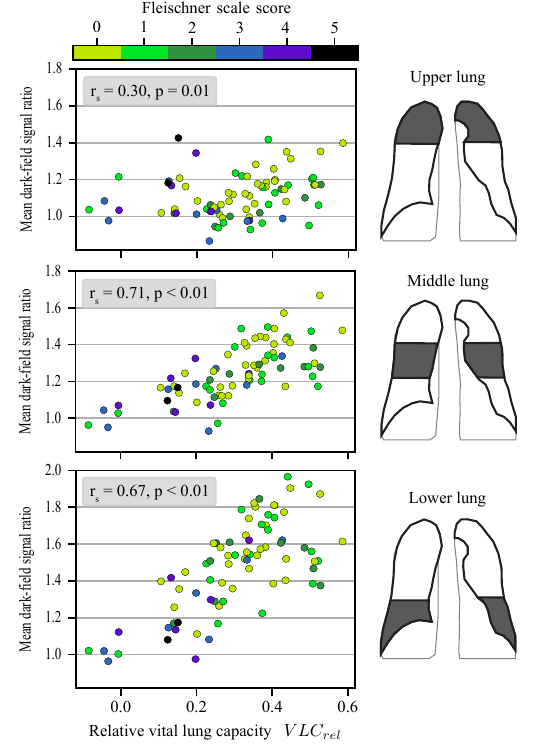}
  \caption{Mean dark-field signal ratio in the three lung regions in relation to $VLC_{\mathrm{rel}}$. Spearman correlation is indicated in the grey boxes. The color map indicates the Fleischner scale score of the study participants, representing the COPD (emphysema) severity (0 - absent, 1 - trace, 2~-~mild, 3 - moderate, 4 - confluent, and 5 - advanced destructive).}
  \label{fig: Mean_DF_signal_change}
\end{figure}

Since in some cases, a nonlinear course was observed in the CC dark-field signal ratio projection graphs, see e.g.\ Fig.~\ref{fig: Reg_examples}\textbf{a}, which a linear fit cannot capture adequately, we evaluated an alternative approach to characterize the regional dark-field signal changes. 
This is visualized in Fig.~\ref{fig: Mean_DF_signal_change} where we calculated the mean dark-field signal ratio in the upper, middle, and lower lung region and evaluated these concerning $VLC_{\mathrm{rel}}$ for 77 study participants.
The other study participants were excluded from this evaluation due to missing CT scans, out-of-image lung regions, or non-standard emphysema type (lymphangioleiomyomatosis, pulmonary fibrosis).

In general, different characteristics of the dark-field signal change in connection with $VLC_{\mathrm{rel}}$ were found in the three lung fields.
There is a weak correlation ($r_{s,\mathrm{upper}}=0.30$, $p=0.01$) of the mean dark-field signal ratio within the upper lung region, but strong correlations within the middle ($r_{s,\mathrm{middle}}=0.71$, $p<0.01$) and lower ($r_{s,\mathrm{lower}}=0.67$, $p<0.01$) lung region. 
In addition to the stronger correlations, the ratio changes over an increasingly broader range in the middle and lower lung fields, respectively.
Combined, the observed characteristics in connection with $VLC_{\mathrm{rel}}$ indicate that this analysis approach allows to capture the different relative alveolar density changes between expiration and inspiration in the distinct lung fields. This is because we expect increasing local lung volume changes in the CC direction due to the dominant motion of the diaphragm during the breathing cycle. The rationale behind this is that we expect an increase in the dark-field signal during exhalation in contracting lung areas due to increased X-ray scattering as a result of the larger number of tissue-air interfaces along the beam path when the alveolar density is increased in these regions.

To evaluate the relationship between $VLC_{\mathrm{rel}}$, signal ratio, and COPD severity, the Fleischner scale score is color-coded within Fig.~\ref{fig: Mean_DF_signal_change}.
This way, it can be observed that the more severe emphysema cases are clustered at small $VLC_{\mathrm{rel}}$ and correspondingly low dark-field signal ratios.
This was to be expected, as emphysematous lung tissue restricts the flexibility of the lungs, accordingly leading to breathing impairments. 
However, data points from low Fleischner score study participants with $VLC_{\mathrm{rel}}$ close to zero and data instances from study participants with more severe emphysema with greater $VLC_{\mathrm{rel}}$ can also be observed.
Moreover, it can be presumed that the study participants with $VLC_{\mathrm{rel}}<0$ did not correctly follow or understand the exhale-inhale breathing instructions as such values are implausible. 
It is therefore important to note that it is crucial how well the study participant follows the breathing instructions during the image recordings. This in turn complicates the analysis of specific relationships between the dark-field signal change and the lung disease severity.
Yet, it may be of interest for further studies to evaluate quantitative differences to explore the possibility of differentiating classes of study participants via the dark-field signal ratios.

\section{Conclusion}
In this study, we presented a framework specifically tailored for the registration of dark-field chest radiographs acquired in different breathing states. 
Apart from the scenario with insufficient transformation regularization, the examined parameter configurations performed well in terms of the evaluation metrics used. 
The selected framework configuration allows to establish an adequate spatial correspondence between dark-field chest radiographs in different respiratory states and thus to perform a local signal change analysis.
This allows us to add a dynamic comparative dimension to the previous static analysis approaches of the dark-field lung signal that considered one breathing state only. 

In addition to the discussed benefits regarding bone and soft-tissue overlap contributions in the lung region, our comparison between the dark-field and attenuation domains using the gradient of the CC axis projections of the registered image signal ratios presented evidence of slope-sensitivity advantages concerning $VLC_{\mathrm{rel}}$ in the dark-field domain. 
This indicates that radiography-based lung function assessment approaches can benefit from the utilization of lung signal information in the dark-field domain in addition to the commonly used attenuation images.
The enhanced lung sensitivity may also help to reduce the number of images that need to be acquired to be able to analyze the signal changes and thus further reduce the radiation dose for patients compared to the standard DCR. 

Our alternative analysis approach of examining the mean signal ratio in the three lung fields provided insight into the distinct behavior of the dark-field signal changes with the breathing capacity. 
The signal characteristics are in good agreement with the expected lung volume changes in the respective lung regions, indicating that our analysis approach is promising to investigate local alveolar density changes during the breathing cycle. This opens up new options for low-dose and rapid lung ventilation assessment via dark-field chest radiography that could improve lung diagnostics considerably.


\begin{thebibliography}{00}


\bibitem{b3} F. Pfeiffer \emph{et al.}, ``Hard-X-ray dark-field imaging using a grating interferometer,'' \emph{Nat. Mater.}, vol. 7, no. 2, pp.~134--137, Feb. 2008,  DOI: 10.1038/nmat2096

\bibitem{yashiro2010origin} W. Yashiro, Y. Terui, K. Kawabata, and A. Momose, ``On the origin of visibility contrast in x-ray Talbot interferometry,'' \emph{Opt. Express}, vol. 18, no. 16, pp.~16890--16901, Aug. 2010, DOI: 10.1364/OE.18.016890

\bibitem{b4} R. Gradl \emph{et al.}, ``Dynamic in vivo chest x-ray dark-field imaging in mice,'' \emph{IEEE Trans. Med. Imaging}, vol. 38, no. 2, pp.~649--656, Feb. 2019, DOI: 10.1109/TMI.2018.2868999

\bibitem{b5} M. Frank \emph{et al.}, ``Dark-field chest X-ray imaging for the assessment of COVID-19-pneumonia,'' \emph{Commun. Med. (Lond.)}, vol. 2, no. 1, pp.~147--156, Nov. 2022, DOI: 10.1038/s43856-022-00215-3

\bibitem{b6} K. Willer \emph{et al.}, ``X-ray dark-field chest imaging for detection and quantification of emphysema in patients with chronic obstructive pulmonary disease: A diagnostic accuracy study,'' \emph{Lancet Digit. Health}, vol. 3, no. 11, pp.~e733--e744, Nov. 2021, DOI: 10.1016/S2589-7500(21)00146-1

\bibitem{b7} T. Urban \emph{et al.}, ``Dark-field chest radiography outperforms conventional chest radiography for the diagnosis and staging of pulmonary emphysema,'' \emph{Invest. Radiol.}, vol. 58, no. 11, pp.~775--781, Nov. 2023, DOI: 10.1097/RLI.0000000000000989

\bibitem{b13} A. Hata \emph{et al.}, ``Dynamic chest X-Ray using a flat-panel detector system: Technique and applications,'' \emph{Korean J. Radiol.}, vol. 22, no. 4, pp.~634--651, Apr. 2021, DOI: 10.3348/kjr.2020.1136

\bibitem{b8} D. L. G. Hill, P. G. Batchelor, M. Holden, and D. J. Hawkes, ``Medical image registration,'' \emph{Phys. Med. Biol.}, vol. 46, no. 3, pp.~R1--R45, Mar. 2001, DOI: 10.1088/0031-9155/46/3/201

\bibitem{b20} A. Sotiras, C. Davatzikos, and N. Paragios, ``Deformable medical image registration: A survey,'' \emph{IEEE Trans. Med. Imaging}, vol. 32, no. 7, pp.~1153--1190, Jul. 2013, DOI: 10.1109/TMI.2013.2265603

\bibitem{b10} K. Murphy \emph{et al.}, ``Toward automatic regional analysis of pulmonary function using inspiration and expiration thoracic CT,'' \emph{Med. Phys.}, vol. 39, no. 3, pp.~1650--1662, Mar. 2012, DOI: 10.1118/1.3687891

\bibitem{li2003establishing} B. Li, G. E. Christensen, E. A. Hoffman, G. McLennan, and J. M. Reinhardt, ``Establishing a normative atlas of the human lung: intersubject warping and registration of volumetric CT images,'' \emph{Acad. Rad.}, vol. 10, no. 3, pp.~255--265, Mar. 2003, DOI: 10.1016/S1076-6332(03)80099-5

\bibitem{pan2018fast} H. Pan, C. Zhou, Q. Zhu, and D. Zheng, ``A fast registration from 3D CT images to 2D X-ray images,'' in ``\emph{2018 IEEE 3rd ICBDA}.,'' Shanghai, China, 2018, pp. 351--355, DOI: 10.1109/ICBDA.2018.8367706

\bibitem{li2014deformable} M. Li \emph{et al.}, ``Deformable image registration for temporal subtraction of chest radiographs,'' \emph{Int. J. Comput. Assist. Radiol. Surg.}, vol. 9, pp.~513--522, Jul. 2014, DOI: 10.1007/s11548-013-0947-y

\bibitem{stavropoulou2021multichannel} A. Stavropoulou, A. Szmul, E. Chandy, C. Veiga, D. Landau, and J. R. McClelland, ``A multichannel feature-based approach for longitudinal lung CT registration in the presence of radiation induced lung damage,'' \emph{Phys. Med. Biol.}, vol. 66, no. 17, pp.~175020, Aug. 2021, DOI: 10.1088/1361-6560/ac1b1d


\bibitem{b11} Y. Lei \emph{et al.}, ``4D-CT deformable image registration using multiscale unsupervised deep learning,'' \emph{Phys. Med. Biol.}, vol. 65, no. 8, p.~085003, Apr. 2020, DOI: 10.1088/1361-6560/ab79c4

\bibitem{b13.1} F. Fyles, T. S. FitzMaurice, R. E. Robinson, R. Bedi, H. Burhan, and M. J. Walshaw, ``Dynamic chest radiography: A state-of-the-art review,'' \emph{Insights Imaging}, vol. 14, no. 1, p.~107, Jun. 19 2023, DOI: 10.1186/s13244-023-01451-4

\bibitem{ding2010comparison} K. Ding, K. Cao, R. E. Amelon, G. E. Christensen, M. L. Raghavan, and J. M. Reinhardt, ``Comparison of intensity- and Jacobian-based estimates of lung regional ventilation.,'' in ``\emph{Third International Workshop on Pulmonary Image Analysis}.,'' 2010, pp. 49-60

\bibitem{xiao2023deep} H. Xiao \emph{et al.}, ``Deep learning-based lung image registration: A review,'' \emph{Comp. Biol. Med.}, vol. 165, p.~107434, Oct. 2023, DOI: 10.1016/j.compbiomed.2023.107434


\bibitem{b22} Y. Yamada \emph{et al.}, ``Difference in the craniocaudal gradient of the maximum pixel value change rate between chronic obstructive pulmonary disease patients and normal subjects using sub-mGy dynamic chest radiography with a flat panel detector system,'' \emph{Eur. J. Radiol.}, vol. 92, pp.~37--44, Jul. 2017, DOI: 10.1016/j.ejrad.2017.04.016

\bibitem{b23} R. Tanaka \emph{et al.}, ``Detectability of regional lung ventilation with flat-panel detector-based dynamic radiography,'' \emph{J. Digit. Imaging}, vol. 21, no. 1, pp.~109--120, Mar. 2008, DOI: 10.1007/s10278-007-9017-8


\bibitem{b14} F. Pfeiffer, T. Weitkamp, O. Bunk, and C. David, ``Phase retrieval and differential phase-contrast imaging with low-brilliance X-ray sources,'' \emph{Nat. Phys.}, vol. 2, no. 4, pp.~258--261, Mar. 2006, DOI: 10.1038/nphys265

\bibitem{b19} T. Urban, W. Noichl, K. J. Engel, T. Koehler, and F. Pfeiffer, ``Correction for X-Ray scatter and detector crosstalk in dark-field radiography,'' \emph{IEEE Trans. Med. Imaging}, vol. 43, no. 7, pp.~2646--2656, Jul. 2024, DOI: 10.1109/TMI.2024.3374974

\bibitem{b17} W. Noichl \emph{et al.}, ``Correction for mechanical inaccuracies in a scanning Talbot-Lau interferometer,'' \emph{IEEE Trans. Med. Imaging}, vol. 43, no. 1, pp.~28--38, Jan. 2024, DOI: 10.1109/TMI.2023.3288358

\bibitem{b18} R. C. Schick \emph{et al.}, ``Correction of motion artifacts in dark-field radiography of the human chest,'' \emph{IEEE Trans. Med. Imaging}, vol. 41, no. 4, pp.~895--902, Apr. 2022, DOI: 10.1109/TMI.2021.3126492


\bibitem{b26} W. Sun, W. J. Niessen, and S. Klein, ``Hierarchical vs. simultaneous multiresolution strategies for nonrigid image registration,'' in \emph{Biomedical Image Registration. WBIR 2012}, 7359th ed. Berlin, Heidelberg: Springer, 2012, Lecture Notes in Computer Science, pp. 60--69. [Online], Available: https://link.springer.com/chapter/10.1007/978-3-642-31340-0\_7. DOI: 10.1007/978-3-642-31340-0\_7

\bibitem{b31} H. Qiu, C. Qin, A. Schuh, K. Hammernik, and D. Rueckert, ``Learning diffeomorphic and modality-invariant registration using b-splines.,'' in ``MIDL,'' Lübeck, Germany, 2021

\bibitem{b27} M. Modat, P. Daga, M. J. Cardoso, S. Ourselin, G. R. Ridgway, and J. Ashburner, ``Parametric non-rigid registration using a stationary velocity field.,'' in ``\emph{2012 IEEE Workshop on Mathematical Methods in Biomedical Image Analysis}.,'' Breckenridge, CO, USA, 2012, pp. 145-150, DOI: 10.1109/MMBIA.2012.6164745

\bibitem{b29} B. B. Avants, C. L. Epstein, M. Grossman, and J. C. Gee, ``Symmetric diffeomorphic image registration with cross-correlation: Evaluating automated labeling of elderly and neurodegenerative brain,'' \emph{Med. Image Anal.}, vol. 12, no. 1, pp.~26--41, Feb. 2008, DOI: 10.1016/j.media.2007.06.004

\bibitem{b30} D. Rueckert and J. Schnabel, ``Medical image registration,'' in \emph{Biomedical image processing}. Berlin, Heidelberg: Springer, 2010, pp. 131--154, DOI: 10.1007/978-3-642-15816-2\_5.

\bibitem{b28} D. Rueckert, L. I. Sonoda, C. Hayes, D. L. G. Hill, M. O. Leach, and D. J. Hawkes, ``Nonrigid registration using free-form deformations: Application to breast MR images,'' \emph{IEEE Trans. Med. Imaging}, vol. 18, no. 8, pp.~712--721, Aug. 1999, DOI: 10.1109/42.796284


\bibitem{devos2019deep} B. D. de Vos, F. F. Berendsen, M. A. Viergever, H. Sokooti, M. Staring, and I. I\v{s}gum, ``A deep learning framework for unsupervised affine and deformable image registration,'' \emph{Med. Image Anal.}, vol. 52, pp.~128--143, Feb. 2019, DOI: 10.1016/j.media.2018.11.010

\bibitem{park2012concave} J. S. Park and S. J. Oh, ``A new concave hull algorithm and concaveness measure for n-dimensional datasets,'' \emph{J. Inf. Sci. Eng.}, vol. 28, no. 3, pp.~587--600, 2012, DOI: 10.6688/JISE.2012.28.3.10

\bibitem{b21} A. Schuh, H. Qiu, and HeartFlow Research, ``deepali: Image, point set, and surface registration in PyTorch.,'' 2023, DOI: 10.5281/zenodo.8170161

\bibitem{bech2010quantitative} M. Bech, O. Bunk, T. Donath, R. Feidenhans'l, C. David, and F. Pfeiffer, ``Quantitative x-ray dark-field computed tomography,'' \emph{Phys. Med. Biol.}, vol. 55, no. 18, pp.~5529--5539, Sep. 2010, DOI: 10.1088/0031-9155/55/18/017

\bibitem{b32} R. J. Pierce, D. J. Brown, M. Holmes, G. Cumming, and D. M. Denison, ``Estimation of lung volumes from chest radiographs using shape information,'' \emph{Thorax}, vol. 34, no. 6, pp.~726--734, Dec. 1979, DOI: 10.1136/thx.34.6.726






\end{thebibliography}
\end{document}